\documentclass[final]{svjour3}
\usepackage{graphicx}
\usepackage{rotating}
\usepackage{amssymb}
\usepackage{mathptmx}
\usepackage[numbers]{natbib}
\makeatletter
\journalname{Journal of Low Temperature Physics}

\bibpunct{}{}{,}{s}{}{,}

\begin{document}

\newcommand{\hdblarrow}{H\makebox[0.9ex][l]{$\downdownarrows$}-}
\title{Silicon PIN diodes as Neganov-Trofimov-Luke cryogenic light detectors}

\author{X. Defay \and E. Mondragon \and J.-C. Lanfranchi \and A. Langenk\"amper \and A. M\"unster \and W. Potzel \and S. Sch\"onert \and S. Wawoczny \and M. Willers.}

\institute{Excellence Cluster Universe and Physics Department, Technical  University of Munich, 85748 Garching, Germany.
\email{name@email.com}}

\maketitle

\begin{abstract}

Cryogenic rare event searches based on heat and light composite calorimeters have a common need for large area photon detectors with high quantum efficiency, good radiopurity and high sensitivity. By employing the Neganov-Trofimov-Luke effect (NTLE), the phonon signal of particle interactions in a semiconductor absorber operated at cryogenic temperatures can be amplified by drifting the photogenerated electrons and holes in an electric field. 
We present here the last results of a Neganov-Trofimov-Luke effect light detector with an electric field configuration optimized to improve the charge collection within the absorber.


\end{abstract}

\section{Introduction}

Cryogenic scintillation experiments such as direct dark matter search \cite{CRESSTLow}, neutrinoless double beta decay \cite{Artusa} (0$\nu$DBD) and coherent neutrino nucleus scattering \cite{Billard} (CNNS) require light detectors with a large area, single photon sensitivity, and high quantum efficiency. The outcome of such experiments highly depends on the performances of the light detectors. The idea of enhancing the signal-to-noise ratio of cryogenic calorimeters to detect optical photons using the NTLE\cite{NeganovTrofimov,Luke,MauriceNTLE2000,MauriceFancy} was initially proposed by Stark et al \cite{Stark2005}. Standard calorimetric light detectors measure the phonon signal induced by the interaction of photons in an absorber. The NTLE can be used to enhance the signal-to-noise ratio of the phonon signal. It is based on the amplification of the phonon signal by drifting the photogenerated electron-hole pairs within an electric field applied in a semiconductor absorber. This effect, very similar to the Joule effect (emission of phonons as electrons traverse through a semiconductor lattice), does not in principle, involve free charge carriers but only photogenerated ones. The detectors investigated in this work are typically operated at temperatures of $\sim$ 25 mK which is well below the Silicon carrier freeze-out temperature ($\sim$40 K). This implies the lack of thermally generated free carriers. Such a photodetector has so far never been used by any large-scale experiment because of problems with charge collection efficiency and reproducibility of such devices.


We discuss here a novel approach taking advantage of a commercially available device with implanted contacts in order to enhance the charge collection and to benefit from the scalability and reliability of an industrial technology. Preliminary results were already presented emphasising that such a detector do not show any sign of degradation of the charge collection in time and present an homogeneous response on the whole surface of the detector\cite{LTDus2015}. We discuss here more thoroughly the NTLE in a photodiode and the analysis of an independent dataset including both thermal and ionization measurements.

\section{A new detector}
\begin{figure}
\begin{center}
\includegraphics[width=0.49\linewidth,keepaspectratio]{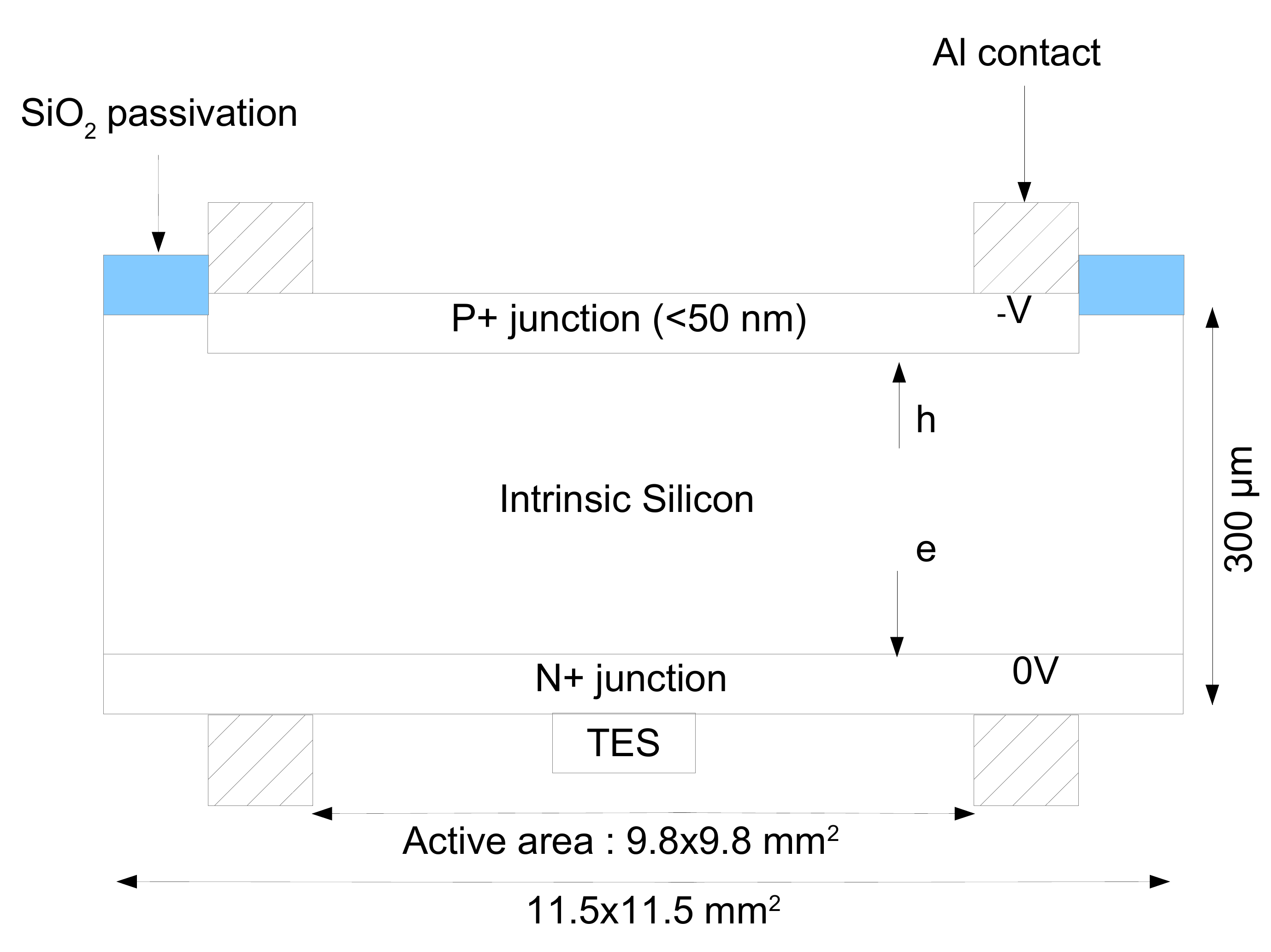}
\includegraphics[width=0.49\linewidth,keepaspectratio]{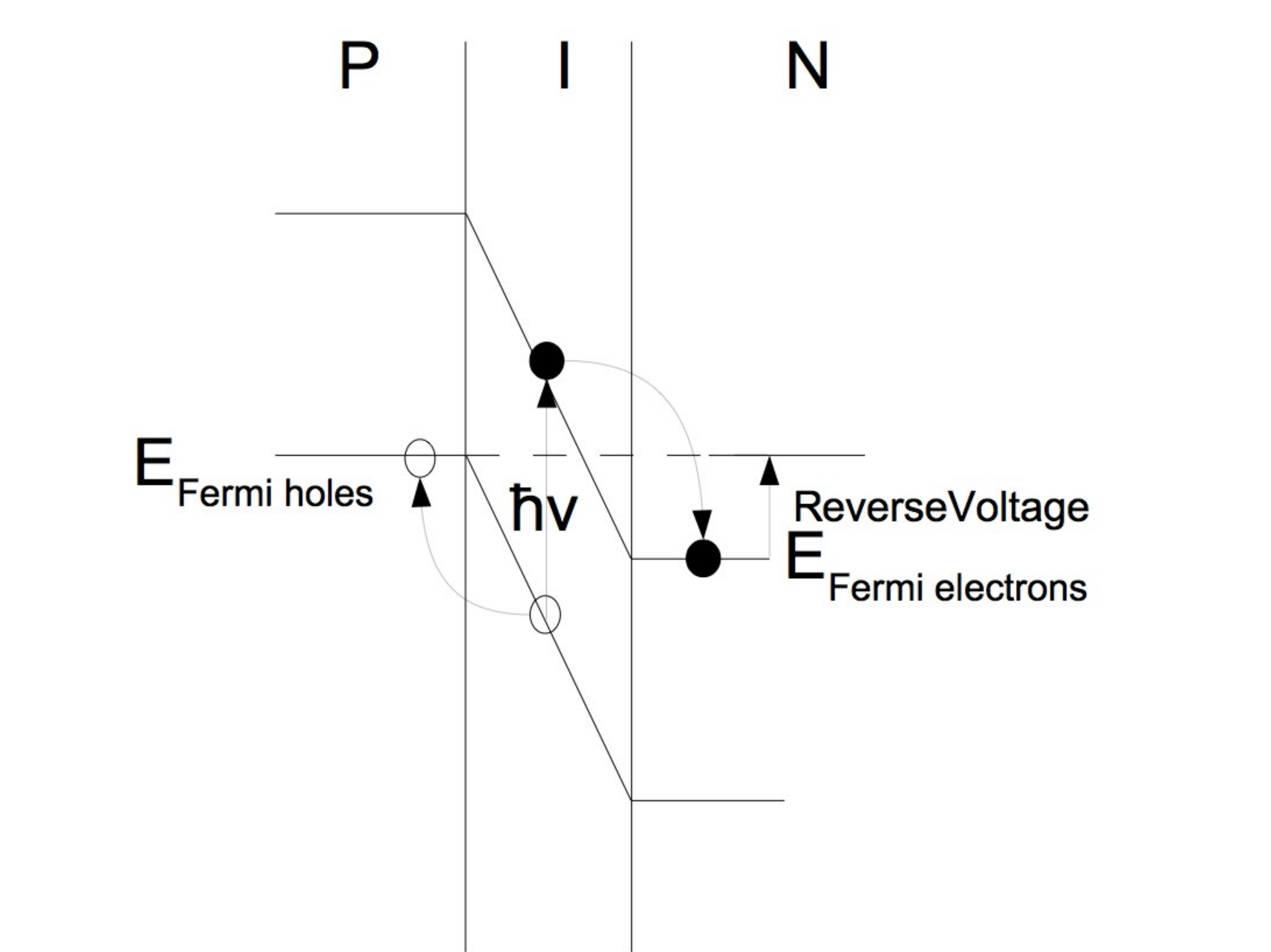}
\end{center}
\caption{{\it Left :} Scheme of the structure of the NTLE detector. The diode is implanted on both sides to form a N+ and a P+ junction. A passivation layer is applied to reduce leakage currents. A TES is glued on the N+ junction to measure the phonon signals. Holes (h) and electrons (e) drift towards the P+ and N+ junctions, respectively. {\it Right :} Band diagram of a PIN photodiode structure with a reverse bias. }
\label{fig1}
\end{figure}

NTLE detectors are traditionally produced by depositing aluminum electrodes onto a semiconductor (Ge or Si). This presents the advantage of simplicity but the disadvantage of drifting carriers along free surfaces (i.e., area of semiconductor without aluminum) where the trapping cross sections are high. The so-called degradation of a detector corresponds to a decrease of the charge collection with time because of charge trapping. Once trapped, the carriers induce a counter electric field opposing the initial field created by the electrodes. An excellent charge collection is therefore required to prevent any degradation and to ensure a constant and high phonon gain. In order to enhance the charge collection, we propose here to use a photodiode as an absorber, with a planar geometry and electrodes with implanted contacts (figure~\ref{fig1} left). The device consists of a  300 $\mu$m thick nearly intrinsic n-type region between N+ and P+ contacts. The P+ contact constitutes the entrance window. The concentration of the dopants in the contacts is high enough ($>10^{19}$ cm$^{-3}$) so that the electrodes remain conductive at very low temperature and the implantation of the entrance window is ultra-shallow ($<$ 50 nm) to minimize the absorption of photons. The resistivity of the intrinsic region is 5.6 k$\Omega$.cm at room temperature. The 11.5x11.5 mm$^{2}$ device is suspended by clamps made of Sintimid which is a thermally and electrically isolating polymer. The phonon signal from the detector is measured with a Transition Edge Sensor (TES) glued onto the N+ junction and read out with a SQUID (Superconducting QUantum Interference Device). The thermal connection to the cryostat consists of a thin gold wire bond between the TES and the thermal bath and the electrical connections are made with aluminum bonds. The charge signal can also be measured with a charge amplifier which, although featuring a much degraded resolution compared with the phonon signal, provides complementary information on the charge collection. Photodiodes are typically fitted with a metal coating on the N+ face. However, a large aluminum layer can decrease the collection of the athermal phonons on the TES. We, therefore, decided to limit the surface of the aluminum contact to a thin frame on both sides of the photodiode. These photodiodes, specially fitted for our experiments, were supplied by Canberra. 

This approach presents several advantages: the free surfaces are limited to a minute area, the electric field within the bulk of the detector can be very high (typically 3000 $\mathrm{V \cdot cm^{-1}}$) and the full depletion of the bulk drastically reduces leakage currents. When the diode is reverse biased, one electron-hole pair is created for every photon reaching the intrinsic region. The high collection efficiency is due to the electric field which separates the carriers on a timescale much shorter than the recombination time. Consequently, the charge collection efficiency in the intrinsic layer reaches 100\% when the photodiode is depleted. The photons to detect need to go through an entrance window and one has to take into account the absorption of the photons within the dead layer. However, the absorption within the entrance window is limited due to the ultra-shallow implanted contact technology. The highly doped implanted contacts imply a high density of free carriers in the region of the contacts which represent a high heat capacity, strongly decreasing the thermal signal. However, the sensitivity remains high because the TES mainly collects athermal phonons.  

The signal to detect consists of a flash of 1 to several thousands of scintillation photons of a wavelength of 430 nm from a nearby scintillating crystal. The photodiodes we used do not have any anti-reflecting coating and the probability of a photon to be reflected is $\sim$44$\%$. This limits the external quantum efficiency since a photon reflected remains undetectable. The non-reflected photons enter the photodiode through the P+ side and their penetration depth follow a Beer-Lambert's law with an absorption depth of $\sim$250 nm. A fraction of the photons is therefore absorbed in the dead layer and the rest is absorbed in the intrinsic region of the photodiode. Considering the penetration depth, the probability of a 430 nm photon to go beyond the intrinsic part is negligible. The supplier estimates the transmission T of the dead layer of 430 nm photons to T=0.76 at room temperature.  The literature remains sparse on the subject of the absorption of photons in doped silicon layers at cryogenic temperature. In addition, we do not know the exact implantation profile of the device. However, we expect that the transmission should slightly increase at low temperature \cite{Jellison,Jellison2}.

\section{The NTLE applied to a photodiode}

Let us consider the case of a photon absorbed within the intrinsic region, when the photodiode is unbiased. Because of the internal electric field of the diode, a photogenerated electron will go to the N+ region, dissipate energy by electron-phonon coupling and end-up at an energy level equal to the Fermi level. Symmetrically, the hole will migrate to the P+ region, also dissipate energy until it reaches the Fermi level. In the absence of trapping the energy dedicated to the creation of the electron-hole pair is recovered when the carriers recombine on the electrodes. The total energy dissipated within the photodiode by scattering of the charges is equal to $\hbar$$\nu$. An electron also flows through the external circuit but at 0 V which implies that no energy was taken from the battery. When a voltage ($V$) is applied to the photodiode (figure~\ref{fig1} right), the electron will scatter down to the N+ region until it ends up at the electron quasi-Fermi-level. The hole will similarly scatter to the P+ region and reach the hole quasi-Fermi-level. This time the total energy dissipated is equal to the initial energy of the photon plus a term which corresponds to the dissipation of the power supplied within the photodiode: $E = \hbar$$\nu + eV$ (with $e$: charge of the electron). 

If we take into account the absorption of the photons within the entrance window, the total phonon gain ($G$) at a given voltage resulting from a photo-absorption of a flash of photons is described by the following equation: 

\begin{equation}
G = \frac{E(V)} {E(0)} = 1+\frac{e \cdot T \cdot  CE(V) \cdot V}{\hbar\nu} ,
\label{eq:1}
\end{equation}

where  \(T \) is the transmission of the P+ junction and \(CE(V) \) a term taking into account the dependence of the collection efficiency on the voltage applied and on the wavelength of the photons detected. The former term, related to the depletion of the photodiode is discussed in the next section. In this simplistic model, we consider that the carriers are either properly collected or recombine directly in the dead layer.

\section{Charge collection}
\label{QE}
An electron-hole pair created in the intrinsic region will be drifted toward the doped regions. Given the thickness of the device and the electric field applied within the intrinsic region, we consider the trapping of carriers in the intrinsic region as negligible. However, in the case of the absorption of a photon within the dead layer, the recombination rate of electron-hole pair is high. A lower drift field is present there, and carriers can only diffuse to the intrinsic region or recombine in the dead layer \cite{dead}. The proportion of carriers created in the dead layer and collected in the intrinsic region increases with the voltage applied. 
    
\begin{figure}
\begin{center}
\includegraphics[width=0.49\linewidth,keepaspectratio]{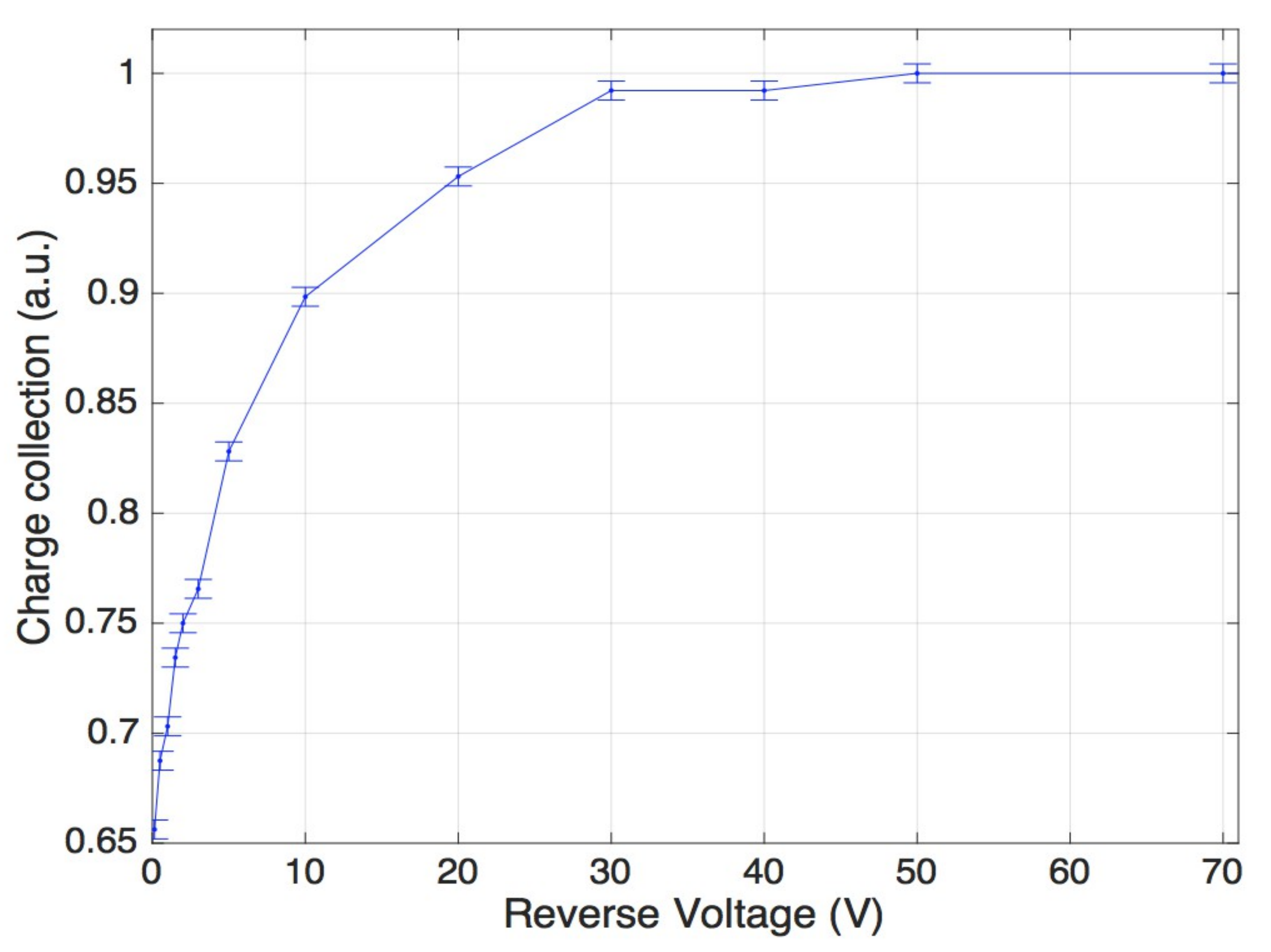}
\includegraphics[width=0.49\linewidth,keepaspectratio]{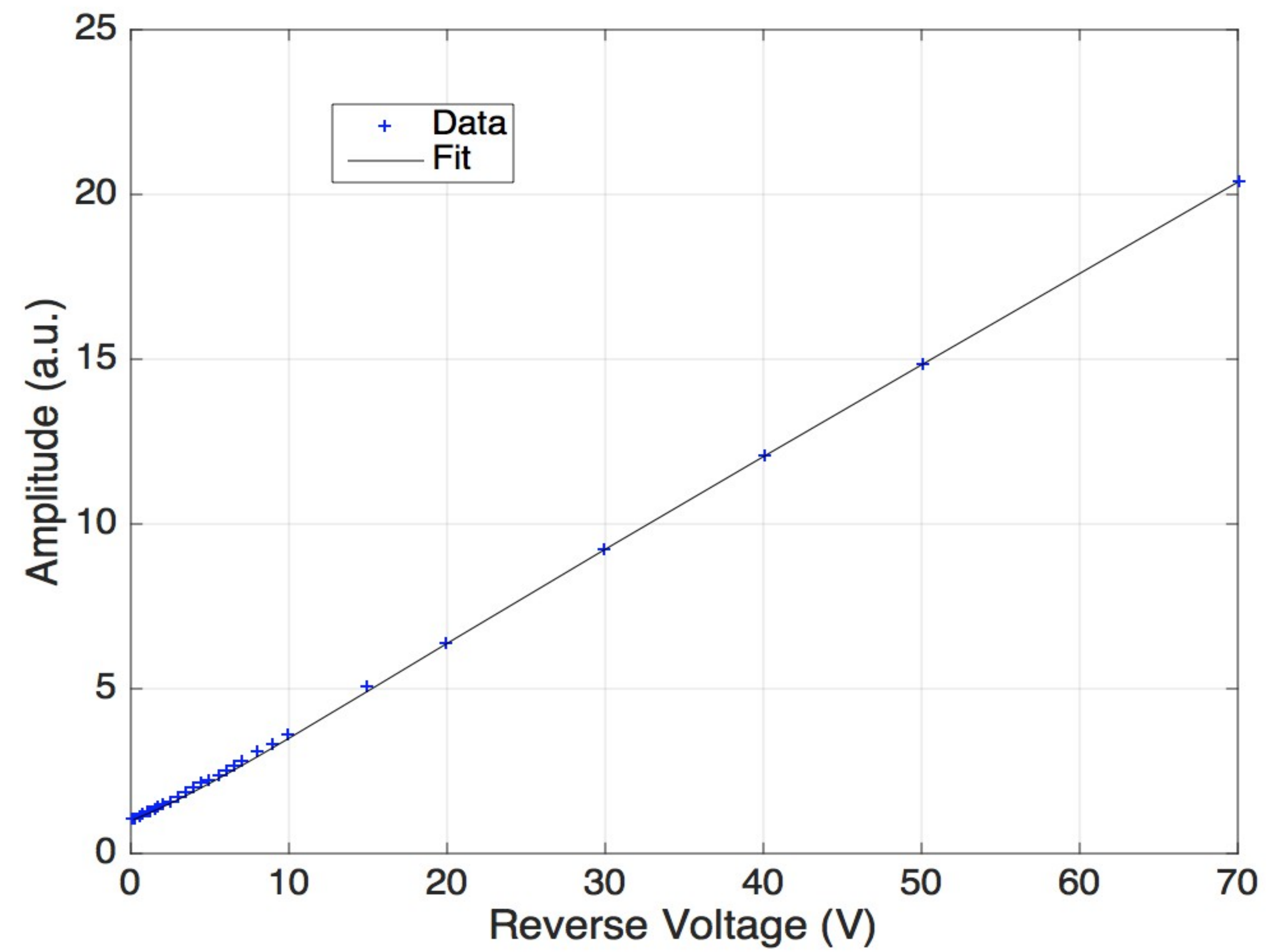}
\end{center}
\caption{{\it Left :} Collection efficiency as a function of the voltage applied to the photodiode at 25 mK for 430 nm photon flashes. The error bar indicate the 1$\sigma$ statistical fluctuations. {\it Right :} Amplitude of the thermal signal plotted as a function of the NTLE voltage for 430 nm photon flashes (the crosses are the experimental values and the best fit using equation \ref{eq:1} is represented by a solid line). The error bars are smaller than the symbols used.}
\label{fig2}
\end{figure}

The amplitude of the charge signals measured with a charge amplifier was studied at low temperature ($\sim$ 25 mK) for photons of 430 nm (see figure~\ref{fig2} left). The photons were emitted by a LED outside the cryostat and guided inside by an optical fiber. Every single event consists of a signal of $\sim 1000$ photons simultaneously absorbed in the detector. The photodiode is able to collect even at 0 V because of the junction potential of the diode. The collection efficiency increases with the voltage until reaching the plateau of maximal collection efficiency. 

\section{Thermal signals}

Figure~\ref{fig2} right represents the evolution of the phonon signals as a function of the voltage. For a voltage higher that 20 V the amplitude follows a pure linear trend. However, for a voltage between 1 V and 20 V, the amplitude of the thermal signals is slightly higher than what is expected according to the simple model previously described by equation \ref{eq:1}. The maximum of the discrepancy reaches $5\%$ at 4 V. We currently do not understand the origin of this effect but conjectured that it could be due to fields induced by charges localized close to the entrance window, creating a potential barrier which is overcome above 20 V. 

The experimental values are fitted using equation \ref{eq:1} while selecting the data when the photodiode is depleted (Reverse Voltage $>$ 20 V). The values of QC(V) are interpolated from the data presented in section \ref{QE}. The slope of the amplitude of the signal allows us to estimate the ratio $\frac{T}{\hbar\nu}$. If we fix $\hbar\nu$= 2.883 eV (the energy of the photons detected), the transmission of the entrance window at very low temperature is T=0.797 $\pm$0.010. This implies that once depleted, the photodiode features a high collection efficiency and that the phonon gain generally behaves as expected. 

We also studied the behavior of the photodiode while biasing it in the forward mode. We found that the diode can withstand a forward bias of about 9 V without drawing any current. Just like for the reverse bias, a gain proportional to the voltage applied was associated with the phonon pulses. This unexpected result could be explained by the presence of a potential barrier between the impurity band at the N+ region and the conduction band at the intrinsic region. In the same way, this potential barrier would exist between the impurity band at the P+ region and the conduction band at the intrinsic region \cite{Simoen, Coon}.

\section{Performances}

\subsection{Determination of the energy resolution}

The method used to determine the resolution consists of sending flashes of 430 nm photons with a different average number of photons\cite{Isaila2012}. The amplitude $x$ of the signal is proportional to the number ($N$) of photons $x=a \cdot N$ ($a$ being a scaling factor). The fluctuations of the amplitude of the signal are the sum of two contributions: the Poisson fluctuations of the number of photons $\sigma _{ph}=a \cdot \sqrt{N}$ which is energy dependent and an energy independent term $\sigma _{0}$. These total fluctuations of the amplitude of the signal can, therefore, be expressed as: 

\begin{equation}
\sigma _{tot}=\sqrt{\sigma _{ph}^2+\sigma _{0}^2}=\sqrt{a \cdot x+\sigma _{0}^2},
\label{eq:2}
\end{equation}

\begin{figure}[htbp]
\begin{center}
\includegraphics[trim = 12mm 65mm 20mm 72mm, clip, width=6cm]{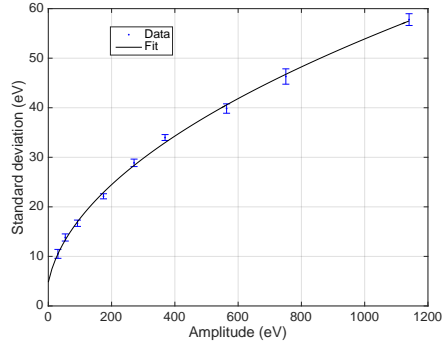}
\end{center}
\caption{$1\sigma$ fluctuations of the amplitude as a function of the amplitude of the phonon signal when changing the average number of photons in each pulse.}
\label{fig3}
\end{figure}

$\sigma _{tot}$ is plotted as a function of the amplitude in figure~\ref{fig3} for an applied voltage of 97 V. The fit using equation \ref{eq:2} gives $\sigma _{0}=5.0 \pm1.3$ eV (which corresponds to 1.75 electron-hole pairs).

The detector was cross-calibrated using the K$_{\alpha}$ line of a $\mathrm{^{55}Fe}$ X-ray source for a voltage applied of 97 V \cite{LTDus2015}. This method yields a baseline noise $\sigma _{0} $ = 5.33$\pm$0.41 eV which is consistent with the result of the 430 nm photon calibration of figure \ref{fig3}.

In order to limit ground loops, the voltage is applied with a battery (97 V corresponds to the highest voltage available). The baseline noise remains typically independent of the voltage applied. However, while applying higher voltages with a generator, we noticed a large increase of the baseline noise above $\sim$200 V probably because of the emergence of leakage currents.

\section{Conclusion and outlook}

We presented here the results of a novel type of NTLE light detector aiming at the improvement of the charge collection by taking advantage of implanted contacts industrially produced. The working principle is experimentally confirmed and once the photodiode is depleted, the thermal gain behaves as expected. The measured baseline noise was evaluated as $\sigma _{0}=5.0\pm1.3$ eV. This encouraging result is still a factor of $\sim$4 larger than the resolution required for the single photon detection at 430 nm. The next developments will concentrate on understanding and optimizing the influence of the implanted contacts on the phonon collection. The implementation of an anti-reflecting coating is also foreseen in order to increase the quantum efficiency. We focused here on the description of a light detector. However, the results obtained with implanted contacts imply that similar detectors could be appropriate as the main absorber for the direct detection of rare events at low energies such as neutrinos (CNNS) or dark matter particles. We also consider in this respect the development of thicker detectors based on the same principle.

\begin{acknowledgements}
This research was supported by the DFG cluster of excellence Origin and Structure of the Universe (www.universe-cluster.de).
\end{acknowledgements}

\pagebreak

\end{document}